\documentclass[journal]{IEEEtran}
\usepackage{cite}
\ifCLASSINFOpdf

\else

   \usepackage[dvips]{graphicx}

\fi
\usepackage{amsmath}
\usepackage{array}
\ifCLASSOPTIONcompsoc
  \usepackage[caption=false,font=normalsize,labelfont=sf,textfont=sf]{subfig}
\else
  \usepackage[caption=false,font=footnotesize]{subfig}
\fi
 \usepackage{multirow}
 \usepackage[normalem]{ulem}
 \useunder{\uline}{\ul}{}
 \usepackage{lscape}
\usepackage{lipsum}
\usepackage{eso-pic}
\usepackage{mathtools}
\usepackage{cuted}

\begin{document}
\title{On the Integral and Derivative Identities of Bivariate Fox's H-Function: Application in Wireless System Performance Analysis}

\author{Puspraj Singh Chauhan$^1$, *Sandeep Kumar$^2$, Imran Shafique Ansari$^3$, 
	
	\thanks{$^1$Department of Electronics $\&$ Communication, Pranveer Singh Institute of Technology, Kanpur, India}
	
	\thanks{$^2$Central Research Laboratory, Bharat Electronics Limited, Gaziabad, India}
	
	\thanks{$^3$James Watt School of Engineering,University of Glasgow, United Kingdom}

	\thanks{}}

\markboth{}
{}


\maketitle

\begin{abstract}
The present work proposes analytical solutions for the integral of bivariate fox's H-function in combination with algebraic, exponential, and complementary error functions. In addition, the work also presents the derivative identities with respect to function arguments. Further, the suitability of the proposed mathematical solutions is verified with reference to wireless communication environment, where a fading behaviour of the channel acquired the bivariate Fox's H-function structure. Further more, asymptotic results for the outage probability and average symbol error probability are presented utilizing the origin probability density function based approach. The obtained results are free from complex analytical functions. At last, the analytical findings of the paper are compared with the numerical results and also with the Monte-Carlo simulation results to confirm their accuracy.
\end{abstract}


\IEEEpeerreviewmaketitle

\section{Introduction}
Bivariate mathematical functions have already demonstrated their importance in the field of communication engineering, as distinct performance metrics evaluation attains analytical structure in bivariate forms \cite{BR1,BR2,BR3,BR4,BR5,BR6,BR7}. In this regard, a wireless environment is considered wherein surroundings of the receiver (Rx) are accompanied by buildings and other sources that impart blockage to direct propagation of signal between the transmitter (Tx) and the receiver (Rx), resulting in simultaneous existence of both multipath and shadowing. Several generalized distributions have been proposed in the literature to model the behaviour of these multipath-shadowed fading scenarios. 
Two fading models namely $\kappa$-$\mu$/inverse Gamma (I-Gamma) and $\eta$-$\mu$ /I-Gamma have been proposed in \cite{BR8}and \cite{BR9} to model the shadowed line-of-sight (LOS) and shadowed non-LOS (NLOS) scenarios, respectively. The utility of $\kappa$-$\mu$/I-Gamma and $\eta$-$\mu$ /I-Gamma fading models in body-centric communications susceptible to the shadowing effect is validated with the measurement data. Although the fading models proposed in \cite{BR8,BR9} are generalized models and find applications in the signal characterization of wearable, vehicular as well as cellular networks etc., but the effect of non-linearity, which is an important factor for 5G and beyond networks, has not been considered in these models.

Later, in \cite{F}, Badarneh has proposed the novel expressions for fading distribution encountering channel non-linearity, i.e. $\alpha$-$\eta$-$\mathcal{F}$ and  $\alpha$-$\kappa$-$\mathcal{F}$. The methodology adopted for deriving the statistics of earlier distributions is as follows: First, the author evaluate the probability density function (PDF) for channels named as $\eta$-$\mathcal{F}$ and $\kappa$-$\mathcal{F}$, respectively. Where multipath behaviour is characterised by $\eta$-$\mu$ \& $\kappa$-$\mu$ and shadowing through inverse-(Nakagami-$m$), followed by  variable transformation approach for evaluating composite PDFs. However, the statistics of proposed distributions do not carry complex functions, still closed-form solutions for other performance metrics are achieved due to laid inherent complexity in the function argument. The authenticity of the earlier statement is verified from cumulative distribution function (CDF) expressions that attain infinite series forms, also lead to truncation error.
Recently, the $\alpha$-$\eta$-$\mu$/I-Gamma and $\alpha$-$\kappa$-$\mu$/I-Gamma distributions, where the effect of non-linearity is considered, has been proposed in \cite{BR10}. These two fading models are more generalized form of $\kappa$-$\mu$/I-Gamma and $\eta$-$\mu$ /I-Gamma distributions. The fundamental statistics and performance metrics of these models are derived in terms of infinite series representations.
The reason for not getting the closed-form solutions proposed in \cite{BR10} is that the closed-form PDF of these fading models can only be represented in the form of bivariate fox's-H function that should have simple function argument. For getting the closed-form expression of the system performance parameters, we need to find out the integral of bivariate fox's H-function in combination with exponential, logarithmic, complementary error function, etc. Moreover, the derivative identities of the bivariate fox's H-function with respect to function arguments are required for deriving the closed-form expressions of other performance parameters, such as low-power analysis of the effective capacity and the optimized threshold in case of cognitive radio performance. The solution to these integral and derivative identities of the bivariate fox's H-function has not been explored till now.

Statistical representations of different performance metrics of wireless systems demand closed-form analysis. Motivated by these findings and as an attempt to obtain the integral and derivative solutions of the bivariate fox's H-function, we derive the novel results that have not been published till date. The obtain results are tiliz to derive the closed-form solution of the performance metrics of the generalized shadowed fading composite fading channel. The main contributions of this work are summarized as follows:\\
$\bullet$	We derive the identities for integral of bivariate fox's H-function in combination with algebraic, exponential, and complementary error function. The solutions of derivative identities with respect to function arguments are also derived. All the results are novel and have not been published till date.\\
$\bullet$ We also substantiate the use of proposed identities in the field of wireless communications. In this regard, a wireless environment is considered wherein surroundings of the receiver (Rx) are accompanied by buildings and other sources that impart blockage to direct transmission between Transmitter (Tx) and Rx, resulting in the simultaneous existence of both multipath and shadowing. The $\alpha$-$\eta$-$\mu$ and I-Gamma fading models are considered to characterize multipath and shadowing, respectively. The PDF of such fading model acquires bivariate Fox's function structure. \\
$\bullet$	The derived identities are applied to obtain closed-form expressions for the performance metrics like outage probability and average symbol error probability (SEP) (coherent and non-coherent). The average SEP results are derived for both additive white Gaussian (AWG) noise  and additive Laplacian noise.\\
$\bullet$	The generality of the $\alpha$-$\eta$-$\mu$/I-Gamma models is highlighted through reduction to some special cases which coincide with existing well-known distributions.\\
$\bullet$	The versatility of the proposed distribution is also verified through simulation of special cases of the results, in which the results of special cases are corroborated with Monte-Carlo simulations, generated independently
by considering the random variable for each case not by altering the channel parameters of the proposed distribution.

The residuum of the paper is structured as follows: In Section-\ref{Sec: System Model}, first the mathematical model of bivariate fox's H-function is discussed, later integral and derivative identities are proposed. Further, in Section-\ref{Sec: Application} the applicability of the solutions is verified over different metrics of a communication system. A comprehensive discussion on the proposed formulations through simulation results is presented in Section-\ref{Sec: Result}, while the essence of the work can be found in Section-\ref{Sec: Conclusion}.

\section{System Model} \label{Sec: System Model}
The analytical representation of dual variable Fox's H-function is given by \cite{Mathai} and is illustrated as:
\begin{equation}\begin{split}\label{Eq:Fox H}
H\left[^x_y\right]=H^{M_1,N_1;M_2,N_2;M_3,N_3}_{P_1,Q_1;P_2,Q_2;P_3,Q_3}~~~~~~~~~~~~~~~~~~~~~~~~~~~~~~~~~~\\\left[^x_y\Big|^{(a_l;\hat{a}_l;\hat{A}_l)_{1,P_1}(c_l;\hat{c}_l)_{1,P_2} (e_l,\hat{e}_l)_{1,P_3}}_{(b_l;\hat{b}_l;\hat{B}_l)_{1,Q_1}(d_l;\hat{d}_l)_{1,Q_2} (f_l,\hat{f}_l)_{1,Q_3}}\right]~~~~~~~~~~~~~~~~~~~~~~~~~~~\\
=-\dfrac{1}{4\pi^2}\int_{\mathcal{L}_1}\int_{\mathcal{L}_2}\phi(s,t)\hat{\phi}_1(s)\hat{\phi}_2(t)x^sy^t ds dt,~~~~~~~~~~~~~~~~~~
\end{split}
\end{equation}
where $M_1=0$ and $\phi(s,t)$, $\hat{\phi}_1(s),$ and $\hat{\phi}_2(t)$ are defined as
\begin{equation}\label{Eq:Phi_s_t}
\phi(s,t)=\dfrac{\prod\limits_{l=1}^{M_1}(b_l-\hat{b_l}s-\hat{B_l}t)\prod\limits_{l=1}^{N_1}(1-a_l+\hat{a_l}s+\hat{A_l}t)}{\prod\limits_{l=N_1+1}^{P_1}(a_l-\hat{a_l}s-\hat{A_l}t)\prod\limits_{l=M_1+1}^{Q_1}(1-b_l+\hat{b_l}s+\hat{B_l}t)},
\end{equation}
\begin{equation}\label{Eq:Phi_s}
\hat{\phi}_1(s)=\dfrac{\prod\limits_{l=1}^{M_2}(d_l-\hat{d_l}s)\prod\limits_{l=1}^{N_2}(1-c_l+\hat{c_l}s)}{\prod\limits_{l=N_2+1}^{P_2}(c_l-\hat{c_l}s)\prod\limits_{l=M_2+1}^{Q_2}(1-d_l+\hat{d_l}s)},
\end{equation}
\begin{equation}\label{Eq:Phi_t}
\hat{\phi}_2(t)=\dfrac{\prod\limits_{l=1}^{M_3}(f_l-\hat{f_l}t)\prod\limits_{l=1}^{N_3}(1-e_l+\hat{e_l}t)}{\prod\limits_{l=N_3+1}^{P_3}(e_l-\hat{e_l}t)\prod\limits_{l=M_3+1}^{Q_3}(1-f_l+\hat{f_l}t)},
\end{equation}
Further, it is presumed that the integrand poles are simple, and the parameters and their respective coefficients are complex and real \& positive, respectively. The convergence of Fox's H function given in Mellin-Barenes double integral will support the relations stated in \cite[Eqs. (2.61-2.65)]{Mathai}.
Here, in this work the integral and differentiation identity of dual variable Fox's H-function will be taken under considerations, i.e., $x$=$ax$ and $y$=$bx$, and then (\ref{Eq:Fox H}) will be re-expressed as 
\begin{equation}\begin{split}\label{Eq:Fox H_1}
H\left[^{ax}_{bx}\right]=-\dfrac{1}{4\pi^2}\int_{\mathcal{L}_1}\int_{\mathcal{L}_2}\phi(s,t)\hat{\phi_1(s)}\hat{\phi_2(t)}(ax)^s(bx)^t ds dt.~~~~~~~~~~~~~~~~~~
\end{split}
\end{equation}
\subsection{Integral Solutions}
In this subsection, the closed-from statistics for definite and indefinite integrals are evaluated over bivariate fox's H-function in coordination with various mathematical functions.\\

{\bf{Theorem 1:}} \emph{For $x>$0, $a>$0, and $b>$0, the integral solution to (\ref{Eq:Fox H_1}) for definite interval $X>$0 is expressed as}
\begin{equation}\label{Eq:Fox_CDF}
\begin{split}
F(X)=\int\limits_{0}^{X}H\left[^{ax}_{bx}\right]dx=XH^{M_1,N_{1}+1;M_2,N_2;M_3,N_3}_{P_{1}+1,Q_{1}+1;P_2,Q_2;P_3,Q_3}\\
\times \left[^{aX}_{bX}\Big|^{(0;1;1)(a_l;\hat{a}_l;\hat{A}_l)_{1,P_1}(c_l;\hat{c}_l)_{1,P_2} (e_l,\hat{e}_l)_{1,P_3}}_{(b_l;\hat{b}_l;\hat{B}_l)_{1,Q_1}(-1;1;1)(d_l;\hat{d}_l)_{1,Q_2} (f_l,\hat{f}_l)_{1,Q_3}}\right].
\end{split}
\end{equation}
{\emph{Proof:}} The integral of function $H\left[^{ax}_{bx}\right]$ for definite interval $X$ is defined as $F(X)=\int\limits_{0}^{X}H\left[^{ax}_{bx}\right]dx$. Substituting (\ref{Eq:Fox H_1}) into the previous relation yields
\begin{equation}\label{Eq:Fox_CDF_Int}
\begin{split}
F(X)=-\dfrac{1}{4\pi^2}\int_{\mathcal{L}_1}\int_{\mathcal{L}_2}\phi(s,t)\hat{\phi_1(s)}\hat{\phi_2(t)}(a)^s(b)^t\\
\times\underbrace{\int\limits_{0}^{X}x^{s+t}dx}_{I_0} ds dt,~~~~~~~~~~~~~~~
\end{split}
\end{equation}
where inner integral $I_0$ is simplified as $I_0=\dfrac{X^{s+t+1}\Gamma(s+t+1)}{\Gamma(s+t+2)}$.
Substituting $I_0$ result back into (\ref{Eq:Fox_CDF_Int}) and utilizing the relation \cite[Eqs. (2.56-2.57)]{Mathai} yields the closed-form solution as expressed in (\ref{Eq:Fox_CDF}).\\

{\bf{Theorem 2:}} \emph{For $x>$0, $a>$0, $b>$0, and $s>$0, the Laplace-Transform (LT) of (\ref{Eq:Fox H_1}) is expressed as}
\begin{equation}\begin{split}\label{Eq:Fox_H_LT}
\mathcal{L}\{H\left[^{ax}_{bx}\right]\}=\dfrac{1}{\hat{s}}H^{M_1,N_{1}+1;M_2,N_2;M_3,N_3}_{P_1+1,Q_{1};P_2,Q_2;P_3,Q_3}\\
\times \left[^{\frac{a}{\hat{s}}}_{\frac{b}{\hat{s}}}\Big|^{(0;1;1)(a_l;\hat{a}_l;\hat{A}_l)_{1,P_1}(c_l;\hat{c}_l)_{1,P_2} (e_l,\hat{e}_l)_{1,P_3}}_{(b_l;\hat{b}_l;\hat{B}_l)_{1,Q_1}(d_l;\hat{d}_l)_{1,Q_2} (f_l,\hat{f}_l)_{1,Q_3}}\right].
\end{split}
\end{equation}
{\emph{Proof:}} The LT for arbitrary function is evaluated by following the identity $\mathcal{L}\{f(\hat{s})\}=\int\limits_{0}^{\infty}f(x)e^{-\hat{s}x}dx$ \cite[Chapter (17.11)]{TI}. Now, substituting (\ref{Eq:Fox H_1}) in earlier relation, one will obtain
\begin{equation}\label{Eq:Fox_LT_Int}
\begin{split}
\mathcal{L}\{H(\hat{s})\}=-\dfrac{1}{4\pi^2}\int_{\mathcal{L}_1}\int_{\mathcal{L}_2}\phi(s,t)\hat{\phi_1(s)}\hat{\phi_2(t)}(a)^s(b)^t\\
\times\underbrace{\int\limits_{0}^{\infty}x^{(s+t+1)-1}e^{-\hat{s}x}dx}_{I_1} ds dt.~~~~~~~~~~~~~~~
\end{split}
\end{equation}
With the aid of \cite[Eq. (3.381.4)]{TI}, the solution to $I_1$ is deduced as follows
\begin{equation}
I_1=\dfrac{\Gamma(s+t+1)}{\hat{s}^{s+t+1}}.
\end{equation} 
Plugging $I_1$ solution back into (\ref{Eq:Fox_LT_Int}) and using \cite[Eq. (2.56-2.57)]{Mathai} with some mathematical manipulations, (\ref{Eq:Fox_H_LT}) is achieved, thus completes the proof.\\
 Table \ref{table1} presents the integral solution of $H\left[^{ax}_{bx}\right]$ with $g(x)$, i.e.,  $\int\limits_{0}^{\infty}H\left[^{ax}_{bx}\right]g(x)dx$, where $g(x)$ is $\exp({-\mathcal{K}\sqrt{x}})$, $\sqrt{x}\exp({-\mathcal{K}\sqrt{x}})$, and $erfc(\sqrt{rx})$ functions. In order to deduce, the closed-form solutions of the given integrals, we first use integral representation of $H\left[^{ax}_{bx}\right]$ given by (\ref{Eq:Fox H_1}), then substituting the Meijer's G- function representation of earlier defined functions using the relation given by \cite[Eq. (8.4.14.2)]{ISV3}
  \begin{eqnarray}
  \exp({-\mathcal{K}\sqrt{x}})=G^{1,0}_{0,1}\left[\mathcal{K}\sqrt{x}\big|^{-}_{0}\right],\\
  erfc(\sqrt{\mathcal{K}x})=\dfrac{1}{\sqrt{\pi}}G^{2,0}_{1,2}\left[\mathcal{K}x\big|^{~~1}_{0;\frac{1}{2}}\right].
  \end{eqnarray}
Univariate Fox's H representation of earlier functions can be obtained following the identity \cite[Eq. (8.3.2.21)]{ISV3}. Then, inserting the obtained analytical expressions along with the Mellin Transform of Fox's H-function \cite[Eq. (2.8)]{Mathai} in continuation with algebraic manipulations and  using identity \cite[ Eqs. (2.56-2.57)]{Mathai}, we obtain the closed-form solutions for the defined integrals as demonstrated in Table \ref{table1}.

\begin{table*}
\caption{Integral Solution of $H\left[^{ax}_{bx}\right]$ for Multiple Functions.}
\centering
\begin{tabular}{|c|c|c|}  
\hline                       
\centering S. No. &  $g(x)$ & $\int\limits_{0}^{\infty}H\left[^{ax}_{bx}\right]g(x)dx$  \\ [0.5ex] 
\hline                    
1 &$e^{-\mathcal{K}\sqrt{x}}$ & $\dfrac{2}{\mathcal{K}^2}H^{M_{1},N_{1}+1;M_2,N_2;M_3,N_3}_{P_{1}+1,Q_{1};P_2,Q_2;P_3,Q_3}\left[^{\frac{a}{\mathcal{K}^2}}_{\frac{b}{\mathcal{K}^2}}\Big|^{(-1;2;2)(a_l;\hat{a}_l;\hat{A}_l)_{1,P_1}(c_l;\hat{c}_l)_{1,P_2} (e_l,\hat{e}_l)_{1,P_3}}_{(b_l;\hat{b}_l;\hat{B}_l)_{1,Q_1}(d_l;\hat{d}_l)_{1,Q_2} (f_l,\hat{f}_l)_{1,Q_3}}\right]$ \\\hline
2 & $\sqrt{x}e^{-\mathcal{K}\sqrt{x}}$ & $\dfrac{2}{\mathcal{K}^3}H^{M_{1},N_{1}+1;M_2,N_2;M_3,N_3}_{P_{1}+1,Q_{1};P_2,Q_2;P_3,Q_3}\left[^{\frac{a}{\mathcal{K}^3}}_{\frac{b}{\mathcal{K}^3}}\Big|^{(-2;2;2)(a_l;\hat{a}_l;\hat{A}_l)_{1,P_1}(c_l;\hat{c}_l)_{1,P_2} (e_l,\hat{e}_l)_{1,P_3}}_{(b_l;\hat{b}_l;\hat{B}_l)_{1,Q_1}(d_l;\hat{d}_l)_{1,Q_2} (f_l,\hat{f}_l)_{1,Q_3}}\right]$
  \\\hline 
3 & $erfc(\sqrt{\mathcal{K}x})$ & $\dfrac{1}{\mathcal{K}\sqrt{\pi}}H^{M_1,N_{1}+2;M_2,N_2;M_3,N_3}_{P_{1}+2,Q_{1}+1;P_2,Q_2;P_3,Q_3}\left[^{\frac{a}{\mathcal{K}}}_{\frac{b}{\mathcal{K}}}\Big|^{(0;1;1)(-\frac{1}{2};1;1)(a_l;\hat{a}_l;\hat{A}_l)_{1,P_1}(c_l;\hat{c}_l)_{1,P_2} (e_l,\hat{e}_l)_{1,P_3}}_{(b_l;\hat{b}_l;\hat{B}_l)_{1,Q_1} (-1;1;1)(d_l;\hat{d}_l)_{1,Q_2} (f_l,\hat{f}_l)_{1,Q_3}}\right]$
  \\\hline

\end{tabular} 
\label{table1}  
\end{table*}
\subsection{Derivative Solutions}
Here, we have derived the analytical solutions for derivative of bivariate Fox's H-function with respect to function independent variable and constants.

{\bf{Theorem 3:}} \emph{For $x>$0, $a>$0, and $b>$0, the derivative of (\ref{Eq:Fox H_1}) with respect to $x$ is derived as}
\begin{eqnarray}\label{Eq: Fox_H_derivative_1}
\begin{split}
\dfrac{dH\left[^{ax}_{bx}\right]}{dx}=
bH^{M_1,N_{1}+1;M_2,N_2;M_3,N_3}_{P_{1}+1,Q_{1}+1;P_2,Q_2;P_3,Q_3}~~~~~~~~~~~~~~~~~\\
\left[^{ax}_{bx}\Big|^{(-1;1;1)((a_l-\hat{A}_l);\hat{a}_l;\hat{A}_l)_{1,P_1}(c_l;\hat{c}_l)_{1,P_2} ((e_l-\hat{e}_l),\hat{e}_l)_{1,P_3}}_{((b_l-\hat{B}_l);\hat{b}_l;\hat{B}_l)_{1,Q_1}(0;1;1)(d_l;\hat{d}_l)_{1,Q_2} ((f_l-\hat{f}_l),\hat{f}_l)_{1,Q_3}}\right],
\end{split}
\end{eqnarray}
\begin{eqnarray}\label{Eq: Fox_H_derivative_2}
\begin{split}
\dfrac{dH\left[^{ax}_{bx}\right]}{dx}=aH^{M_1,N_{1}+1;M_2,N_2;M_3,N_3}_{P_{1}+1,Q_{1}+1;P_2,Q_2;P_3,Q_3}~~~~~~~~~~~~~~~~~~~~\\
\left[^{ax}_{bx}\Big|^{(-1;1;1)((a_l-\hat{a}_l);\hat{a}_l;\hat{A}_l)_{1,P_1}((c_l-\hat{c}_l);\hat{c}_l)_{1,P_2} (e_l,\hat{e}_l)_{1,P_3}}_{((b_l-\hat{b}_l);\hat{b}_l;\hat{B}_l)_{1,Q_1}(0;1;1)((d_l-\hat{d}_l);\hat{d}_l)_{1,Q_2} (f_l,\hat{f}_l)_{1,Q_3}}\right].
\end{split}
\end{eqnarray}
{\emph{Proof:}} Taking derivative of (\ref{Eq:Fox H_1}) with respect to $x$, it insinuates 
\begin{equation}\label{Eq:Fox_Int_Deriv_1}
\begin{split}
\dfrac{dH\left[^{ax}_{bx}\right]}{dx}=-\dfrac{1}{4\pi^2}\int_{\mathcal{L}_1}\int_{\mathcal{L}_2}\phi(s,t)\hat{\phi_1(s)}\hat{\phi_2(t)}~~~~~~\\\times (s+t)a^sb^tx^{s+t-1}  ds dt.~~~~~~~~~~~~~~~~~
\end{split}
\end{equation}
Now setting $t$=$t'$+$1$ and  performing some mathematical manipulations, we obtain
\begin{equation}\label{Eq:Fox_Int_Deriv_2}
\begin{split}
\dfrac{dH\left[^{ax}_{bx}\right]}{dx}=-\dfrac{1}{4\pi^2}\int_{\mathcal{L}_1}\int_{\mathcal{L}_2}\phi(s,t')\hat{\phi_1(s)}\hat{\phi_2(t')}~~~~~\\
\times(s+t'+1)a^sb^tx^{s+t'} ds dt',~~~~~~~~~~~~
\end{split}
\end{equation}
Applying  \cite[ Eqs. (2.56-2.57)]{Mathai} concludes the proof as expressed in (\ref{Eq: Fox_H_derivative_1}). Similarly, 
(\ref{Eq: Fox_H_derivative_2}) will be obtained by setting $s$=$s'$+$1$ and following the similar procedure as stated above. \\

{\bf{Theorem 4:}} \emph{For $x>$0, $a>$0, and $b>$0, the derivatives of (\ref{Eq:Fox H_1}) with respect to arguments $a$ and $b$ are derived as}
\begin{equation}\label{Eq: Fox_H_derivative_a}
\begin{split}
 \dfrac{dH\left[^{ax}_{bx}\right]}{da}=\dfrac{1}{a}H^{M_1,N_1;M_2,N_2+1;M_3,N_3}_{P_1,Q_1;P_2+1,Q_2+1;P_3,Q_3}~~~~~~~~~~~\\
 \left[^{a}_{b}\Big|^{(a_l;\hat{a}_l;\hat{A}_l)_{1,P_1}(c_l;\hat{c}_l)_{1,P_2}(0;1)(e_l,\hat{e}_l)_{1,P_3}}_{(b_l;\hat{b}_l;\hat{B}_l)_{1,Q_1}(d_l;\hat{d}_l)_{1,Q_2}(1;1) (f_l,\hat{f}_l)_{1,Q_3}}\right],
 \end{split}
\end{equation}
\begin{equation}\label{Eq: Fox_H_derivative_b}
\begin{split}
 \dfrac{dH\left[^{ax}_{bx}\right]}{db}=\dfrac{1}{b}H^{M_1,N_1;M_2,N_2;M_3,N_3+1}_{P_1,Q_1;P_2,Q_2;P_3+1,Q_3+1}~~~~~~~~~~~~~~\\
 \left[^{a}_{b}\Big|^{(a_l;\hat{a}_l;\hat{A}_l)_{1,P_1}(c_l;\hat{c}_l)_{1,P_2}(e_l,\hat{e}_l)_{1,P_3}(0;1)}_{(b_l;\hat{b}_l;\hat{B}_l)_{1,Q_1}(d_l;\hat{d}_l)_{1,Q_2} (f_l,\hat{f}_l)_{1,Q_3}(1;1)}\right].
 \end{split}
\end{equation}
{\emph{Proof:}} Using the integral relation for bivariate Fox's H-function derived in (\ref{Eq:Fox H_1}), differentiating it with respect to function argument $a$ or $b$, and finally making use of \cite[ Eqs. (2.56-2.57)]{Mathai}, the derivative expressions are derived in (\ref{Eq: Fox_H_derivative_a}) and (\ref{Eq: Fox_H_derivative_b}).

 To the author's best knowledge, these derived expressions in (\ref{Eq:Fox_CDF}), (\ref{Eq:Fox_H_LT}),  (\ref{Eq: Fox_H_derivative_1}), (\ref{Eq: Fox_H_derivative_2}), (\ref{Eq: Fox_H_derivative_a}), (\ref{Eq: Fox_H_derivative_b}), and Table-\ref{table1} are novel and have not been reported in literature yet. The convergence of these expression defines if \cite[Eqs. 2.61-2.64]{Mathai} satisfies.
\section{Applications to Wireless Communication}\label{Sec: Application}
The present section verifies the importance of previously derived expressions in the field of digital communication. For this, we have presented the closed-form statistics of different performance metrics of wireless communication system that are highly recommended while studying the system behaviour under fading environments. In this context, the closed-form analytical solutions are provided for the fading environment that have PDF in bivariate Fox's H-function.

The fading scenario that we consider here is composed of correlated surfaces resulting in non-homogeneous/non-linear environment and have a cumulative impact on both multipath and shadowing of the propagating signal. In this context, we have taken generalised $\alpha$-$\eta$-$\mu$ distribution to account for non-linearity and shadowing with inverse-Gamma distribution. In addition, we also derive various system metrics that are used to analyse the system behaviour and allow to understand how well the system performs under given physical conditions.
\subsection{Bivariate Fading Model}
The power PDF for random variable (RV), i.e., $Y$ following $\alpha$-$\eta$-$\mu$ distribution is defined as \cite[Eq. (1)]{Alpha_eta_mu}
\begin{equation}\label{eq:alpha_eta_mu_PDF}
	f_Y(\gamma)=\dfrac{\sqrt{\pi}\mathcal{M}}{\bar{\gamma}^{\frac{\alpha}{2}\hat{\mathcal{{P}}}}}\gamma^{\frac{\alpha}{2}\hat{\mathcal{{P}}}-1}e^{-\frac{\mathcal{Q}\gamma^{\frac{\alpha}{2}}}{\bar{\gamma}^{\frac{\alpha}{2}}}}I_{{\mathcal{P}}}\left(\dfrac{\hat{\mathcal{Q}}\gamma^{\frac{\alpha}{2}}}{\bar{\gamma}^{\frac{\alpha}{2}}}\right),
\end{equation} 
where $\hat{\mathcal{P}}$=$\mu$+$\frac{1}{2}$, ${\mathcal{P}}$=$\mu$-$\frac{1}{2}$, $\mathcal{M}$=$\frac{\alpha \mu^{\hat{\mathcal{P}}}h^{\mu}}{\Gamma(\mu)H^{{\mathcal{P}}}}$, $\mathcal{Q}$=${2\mu h}$, $\hat{\mathcal{Q}}$=${2\mu H}$, $\Gamma(.)$ is the Gamma function, $I_r(.)$ is the $r^{th}$ order modified Bessel function of kind $1^{st}$, $\mu$ signifies the number of multipath clusters, $\alpha$ denotes the propagation medium non-linearity, and $h$ \& $H$ relate to formats I \& II are defined in \cite[Eq. (2) \& Eq. (3)]{Alpha_eta_mu}.
Alternatively, (\ref{eq:alpha_eta_mu_PDF}) can be re-written in Fox's H function with the aid of \cite[Eqs. (1.59) \& (1.60)]{Mathai} and applying some algebraic manipulations as
\begin{equation}\label{eq:alpha_eta_mu_PDF_Fox}
	\begin{split}
		f_Y(\gamma)=\frac{\mathcal{R}}{\bar{\gamma}}H^{1,0}_{0,1}\left[\frac{\mathcal{T}^{\frac{2}{\alpha}}\gamma}{\bar{\gamma}}\Big|^{~~~~~-}_{((\hat{\mathcal{P}}-\frac{2}{\alpha});\frac{2}{\alpha})}\right]
		H^{1,1}_{1,2}\left[\dfrac{\hat{\mathcal{T}}^{\frac{\alpha}{2}} \gamma}{\bar{\gamma}}\Big|^{~~~~(\frac{1}{2};\frac{2}{\alpha})}_{\left({\mathcal{P}};\frac{2}{\alpha}\right),\left(-{\mathcal{P}};\frac{2}{\alpha}\right)}\right],
	\end{split}
\end{equation}
where $\mathcal{R}=\left(\dfrac{2}{\alpha}\right)^2\mathcal{M}\mathcal{T}^{\frac{2}{\alpha}-\hat{P}}$, ${\mathcal{T}}$=${2\mu (h-H)}{}$, and $\hat{\mathcal{T}}$=${4\mu H}{}$.\\
An I-Gamma RV $Z$, having shape parameter $m_s$ and scale parameter $\Omega$ is given as \cite[Eq. (2)]{BR10}
\begin{equation}\label{eq:I_Gamma_PDF_1}
f_Z(z)=\dfrac{\Omega^{m_s}}{\Gamma(m_s)z^{m_s+1}}\exp\left(-\dfrac{\Omega}{z}\right).
\end{equation}
Further in \cite{BR10}, considering $\frac{m_s}{\Omega}$ with unity $\Omega$, the expression (\ref{eq:I_Gamma_PDF_1}) is re-written as
\begin{equation}\label{eq:I_Gamma_PDF}
f_Z(z)=\dfrac{m_s^{m_s}}{\Gamma(m_s)z^{m_s+1}}\exp\left(-\dfrac{m_s}{z}\right).
\end{equation}
{\bf{Theorem 5:}} \emph{For $\alpha>$0, $\mu>$0, 0$<\eta<$1 or $\eta>$1, $m_s>$0, the instantaneous signal-to-noise-ratio (SNR) PDF is given as}
\begin{equation}\label{eq:Composite_PDF_SNR_CF}
\begin{split}
f_Y(\gamma)=\frac{\mathcal{R}}{\Gamma(m_s+1)\bar{\gamma}}~~~~~~~~~~~~~~~~~~~~~~~~~~~~~~~~\\
\times H^{0,1;1,0;1,1}_{1,0;0,1;1,2}\left[^{\frac{\mathcal{T}^{\frac{2}{\alpha}}\gamma}{\bar{\gamma} m_s}}_{\frac{\hat{\mathcal{T}}^{\frac{\alpha}{2}}\gamma}{\bar{\gamma} m_s}}\Bigg|^{(-m_s;1;1)~~~~-~~~~~~~~(\frac{1}{2};\frac{2}{\alpha})}_{~~~~-~((\hat{\mathcal{P}}-\frac{2}{\alpha});\frac{2}{\alpha})~(\mathcal{P};\frac{2}{\alpha})(-\mathcal{P};\frac{2}{\alpha}))}\right].
\end{split}
\end{equation}
{\emph{Proof:}} The power PDF of composite $\alpha$-$\eta$-$\mu$/I-Gamma is obtained through averaging the conditional $\alpha$-$\eta$-$\mu$ fading as a function of mean signal power of shadowed fading, i.e., $Z$, which follows
\begin{equation}\label{eq:Composite_Exact}
f_Y(\gamma)=\int\limits_{0}^{\infty}f_{Y|Z}(\gamma|z)f_Z(z)dz .
\end{equation}
Using the relation for the $\alpha$-$\eta$-$\mu$ PDF as expressed in (\ref{eq:alpha_eta_mu_PDF_Fox}), the conditioned PDF is given as
\begin{equation}\label{eq:Alpha_eta_mu_Conditioned_PDF}
\begin{split}
f_{Y|Z}(\gamma|z)=\frac{\mathcal{R}}{\bar{\gamma}z}H^{1,0}_{0,1}\left[\frac{\mathcal{T}^{\frac{2}{\alpha}}\gamma}{\bar{\gamma}z}\Big|^{~~~~~-}_{((\hat{\mathcal{P}}-\frac{2}{\alpha});\frac{2}{\alpha})}\right]~~~~~~~~~~\\
	\times	H^{1,1}_{1,2}\left[\dfrac{\hat{\mathcal{T}}^{\frac{\alpha}{2}} \gamma}{\bar{\gamma}z}\Big|^{~~~~(\frac{1}{2};\frac{2}{\alpha})}_{\left({\mathcal{P}};\frac{2}{\alpha}\right),\left(-{\mathcal{P}};\frac{2}{\alpha}\right)}\right].~~~~~~~~~~~~
\end{split}
\end{equation}
Putting (\ref{eq:Alpha_eta_mu_Conditioned_PDF}) \& (\ref{eq:I_Gamma_PDF}) into (\ref{eq:Composite_Exact}), with the aid of \cite[Eq. (1.125)]{Mathai} , \cite[Eqs. (1.2) \& (1.3)]{Mathai} in (\ref{eq:alpha_eta_mu_PDF_Fox}), and setting $1/z$=$v$, it insinuates
\begin{equation}\label{eq:Composite_PDF_INT}
\begin{split}
f_Y(\gamma)=\frac{\mathcal{R}m_s^{m_s}}{\Gamma(m_s)\bar{\gamma}}
\dfrac{1}{(2\pi i)^2}\int_{\mathcal{L}_1}\int_{\mathcal{L}_2}\Gamma\left(\left(\hat{\mathcal{P}}-\frac{2}{\alpha}\right)-r\frac{2}{\alpha}\right)~~~~~~\\
\times\dfrac{\Gamma\left(\mathcal{P}-t\frac{2}{\alpha}\right)\Gamma\left(1-\frac{1}{2}+t\frac{2}{\alpha}\right)}{\Gamma\left(1-(-\mathcal{P})+t\frac{2}{\alpha}\right)}\left(\frac{\mathcal{T}^{\frac{2}{\alpha}}\gamma}{\bar{\gamma}}\right)^{r}\left(\dfrac{\hat{\mathcal{T}}^{\frac{\alpha}{2}} \gamma}{\bar{\gamma}}\right)^{t}\\
\times \underbrace{\int\limits_{0}^{\infty}v^{m_s+r+t+1-1}\exp\left(-{m_s v}\right)dv dr dt}_{I_2}.~~~~~~~~~~~~~~~
\end{split}
\end{equation}
The solution to inner integral is deduced with the aid of \cite[Eq. (3.381.4)]{TI}, i.e., $I_2=\dfrac{\Gamma(1+r+t+m_s)}{m_s^{1+r+t+m_s}}$.
Plugging $I_0$ into (\ref{eq:Composite_PDF_INT}), yields
\begin{equation}\label{eq:Composite_PDF_INT1}
\begin{split}
f_Y(\gamma)=\frac{\mathcal{R}}{\Gamma(m_s+1)\bar{\gamma}}
\dfrac{1}{(2\pi i)^2}\int_{\mathcal{L}_1}\int_{\mathcal{L}_2}~~~~~~~~~~~~~~~~~~~~~~~~~~~~~~~~~~~~~~~~~~~~~~~~~~~~~~~~~~~~~~~\\
\times\Gamma(1-(0-m_s)+r+t)\Gamma\left(\left(\hat{\mathcal{P}}-\frac{2}{\alpha}\right)-r\frac{2}{\alpha}\right)~~~~~~~~~~~~~~~~~~~~~~~~~~~~~~~~~~~~~~~~~~~~~~\\
\times\dfrac{\Gamma\left(\mathcal{P}-t\frac{2}{\alpha}\right)\Gamma\left(1-\frac{1}{2}+t\frac{2}{\alpha}\right)}{\Gamma\left(1-(-\mathcal{P})+t\frac{2}{\alpha}\right)}\left(\frac{\mathcal{T}^{\frac{2}{\alpha}}\gamma}{\bar{\gamma}m_s}\right)^{r}\left(\dfrac{\hat{\mathcal{T}}^{\frac{\alpha}{2}} \gamma}{\bar{\gamma}m_s}\right)^{t} dr dt .~~~~~~~~~~~~~~~~~~~~~~~~~~~~~~~~~~~~
\end{split}
\end{equation}
Utilizing \cite[Eqs. (2.56-2.60)]{Mathai}, it concludes the proof as theorized in (\ref{eq:Composite_PDF_SNR_CF}). 

{\bf{Theorem 6:}} The following expression for the generalized MGF over $\alpha$-$\eta$-$\mu$/I-Gamma fading holds
\begin{equation}\label{eq:MGF_CF}
\begin{split}
M_Y^{n}(-s)=\frac{\mathcal{R}}{s^{n+1}\Gamma(m_s+1)\bar{\gamma}}~~~~~~~~~~~~~~~~~~~~~~~~~~~~~~~~\\
\times H^{0,2;1,0;1,1}_{2,0;0,1;1,2}\left[^{\frac{\mathcal{T}^{\frac{2}{\alpha}}\gamma}{s\bar{\gamma} m_s}}_{\frac{\hat{\mathcal{T}}^{\frac{\alpha}{2}}\gamma}{s\bar{\gamma} m_s}}\Bigg|^{(-n;1;1)(-m_s;1;1)~-~~~(\frac{1}{2};\frac{2}{\alpha})}_{~~~~-~((\hat{\mathcal{P}}-\frac{2}{\alpha});\frac{2}{\alpha})~(\mathcal{P};\frac{2}{\alpha})(-\mathcal{P};\frac{2}{\alpha}))}\right].
\end{split}
\end{equation}
{\emph{Proof:}} For a given RV $Y$, the representation of generalized MGF is defined as \cite{MGF}, i.e., $M_Y^{n}(-s)=\int\limits_{0}^{\infty}\gamma^n e^{-s\gamma}f_Y(\gamma)d\gamma$. Now, substituting (\ref{eq:Composite_PDF_INT1}) in the earlier relation (\ref{eq:MGF_CF}), we obtain 
\begin{equation}\label{Eq:MGF_Int}
\begin{split}
M_Y^{n}(-s)=\frac{\mathcal{R}}{\Gamma(m_s+1)\bar{\gamma}}
\dfrac{1}{(2\pi i)^2}\int_{\mathcal{L}_1}\int_{\mathcal{L}_2}~~~~~~~~~~~~~~~\\
\times\Gamma(1-(0-m_s)+r+t)\Gamma\left(\left(\hat{\mathcal{P}}-\frac{2}{\alpha}\right)-r\frac{2}{\alpha}\right)~~~\\
\times\dfrac{\Gamma\left(\mathcal{P}-t\frac{2}{\alpha}\right)\Gamma\left(1-\frac{1}{2}+t\frac{2}{\alpha}\right)}{\Gamma\left(1-(-\mathcal{P})+t\frac{2}{\alpha}\right)}\left(\frac{\mathcal{T}^{\frac{2}{\alpha}}\gamma}{\bar{\gamma}m_s}\right)^{r}\left(\dfrac{\hat{\mathcal{T}}^{\frac{\alpha}{2}} \gamma}{\bar{\gamma}m_s}\right)^{t}\\
\times \int\limits_{0}^{\infty}\gamma^{n+r+t+1-1}e^{-s\gamma}d\gamma dr dt~~~~~~~~~~~~~~~~~~~~~~~~~~
\end{split}
\end{equation}
Subsequently using the relations \cite[Eq. (3.381.4)]{TI} and \cite[Eqs. (2.56-2.60)]{Mathai} with some mathematical manipulations, (\ref{eq:MGF_CF}) is achieved, thus completes the proof.

\subsection{System Performance Metrics}
In this subsection, we evaluate closed-form statistics of physical layer performance metrics such as the outage probability and average SEP. 
\subsubsection{Outage Probability}
Outage probability provides the informative response in terms of the probability that the received instantaneous SNR is below a specified threshold value $\gamma_{th}$. And if the SNR is above the predefined threshold SNR (i.e., $\gamma_{th}$) then the impact of channel on the transmission is minimal. Analytically, the result for outage probability is derived by using the relation $P_{out}(\gamma_{th})=[Y<\gamma_{th}]=F_Y(\gamma_{th})$, and following the steps as mentioned in Theorem-1. The expression for outage probability is expressed as
\begin{equation}
\begin{split}
P_{out}(\gamma_{th})= \frac{\mathcal{R}\gamma_{th}}{\Gamma(m_s+1)\bar{\gamma}}~~~~~~~~~~~~~~~~~~~~~~~~~~~~~~~~\\
\times H^{0,2;1,0;1,1}_{2,1;0,1;1,2}\left[^{\frac{\mathcal{T}^{\frac{2}{\alpha}}\gamma}{\bar{\gamma} m_s}}_{\frac{\hat{\mathcal{T}}^{\frac{\alpha}{2}}\gamma}{\bar{\gamma} m_s}}\Bigg|^{(0,1,1)(-m_s;1;1)~~~~-~~~~~~~~(\frac{1}{2};\frac{2}{\alpha})}_{(-1,1,1)~((\hat{\mathcal{P}}-\frac{2}{\alpha});\frac{2}{\alpha})~(\mathcal{P};\frac{2}{\alpha})(-\mathcal{P};\frac{2}{\alpha}))}\right].
\end{split}
\end{equation}

\subsubsection{Average SEP Analysis}
Average SEP is also considered as one of the important metrics to analyse the quality of signal transmission through wireless channel. The noise characteristics that have been considered are AWG noise and additive Laplacian noise. Here, we have derived the closed-form relations for coherent as well as for non-coherent SEP over different modulation schemes. To obtain the analytical results when channel experiences fading is given as
\begin{equation}\label{Eq:BER_Exact}
\bar{P}(e)=\int\limits_{0}^{\infty}P_e(\gamma)f_Y(\gamma)d\gamma,
\end{equation}
where $P_e(\gamma)$ is the error probability under non-fading environments.\\
\paragraph{Average SEP with AWG Noise} The results for error probability under coherent and non-coherent schemes are derived with AWG noise.\\
{\bf{Corollary 1:}} (Average Coherent SEP). The general expression for average error probability under coherent modulation scheme is given as
\begin{equation}\label{Eq:Coherent_BER_CF}
\begin{split}
\bar{P}_{C}(e)=\dfrac{\mathcal{A}_c\mathcal{R}}{\Gamma(m_s+1)\bar{\gamma}\mathcal{B}_c\sqrt{\pi}}~~~~~~~~~~~\\
\times H^{0,3;1,0;1,1}_{3,1;0,1;1,2}\left[^{\frac{\mathcal{T}^{\frac{2}{\alpha}}}{\bar{\gamma} m_s\mathcal{B}_c}}_{\frac{\hat{\mathcal{T}}^{\frac{\alpha}{2}}}{\bar{\gamma} m_s\mathcal{B}_c}}\Bigg|^{(0;1;1)(-\frac{1}{2};1;1)(-m_s,1,1)-~~~(\frac{1}{2};\frac{2}{\alpha})}_{(-1;1;1)~((\hat{\mathcal{P}}-\frac{2}{\alpha});\frac{2}{\alpha})~(\mathcal{P};\frac{2}{\alpha})(-\mathcal{P};\frac{2}{\alpha}))}\right],
\end{split}
\end{equation}
where constants $A_c$ and $B_c$ for different coherent modulation formats are defined in \cite[Table I]{Alpha_eta_mu}.\\
{\emph{Proof:}} The unified representation of probability of error for non-fading channel under coherent modulation is given as
\begin{equation}\label{Eq:AWGN_Coherent}
P_e(\gamma)=A_c erfc(\sqrt{B_c\gamma}).
\end{equation} 
Now substituting, (\ref{Eq:AWGN_Coherent}) and (\ref{eq:Composite_PDF_SNR_CF}) into (\ref{Eq:BER_Exact}) and following the relation derived Table \ref{table1}, the proof concludes as proposed in (\ref{Eq:Coherent_BER_CF}).\\
{\bf{Corollary 2:}} (Average Non-Coherent SEP). The probability of error expression under non-coherent modulation schemes is presented as
\begin{equation}\label{Eq:BER_Non_Coherent_CF}
\begin{split}
\bar{P}_{NC}(e)=\dfrac{\mathcal{A}_{nc}\mathcal{R}}{\Gamma(m_s+1)\bar{\gamma}\mathcal{B}_{nc}}~~~~~~~~~~~\\
\times H^{0,2;1,0;1,1}_{2,0;0,1;1,2}\left[^{\frac{\mathcal{T}^{\frac{2}{\alpha}}}{\bar{\gamma} m_s\mathcal{B}_{nc}}}_{\frac{\hat{\mathcal{T}}^{\frac{\alpha}{2}}}{\bar{\gamma} m_s\mathcal{B}_{nc}}}\Bigg|^{(0;1;1)(-m_s,1,1)~~~~-~~~~~~~~(\frac{1}{2};\frac{2}{\alpha})}_{~~~-~~~((\hat{\mathcal{P}}-\frac{2}{\alpha});\frac{2}{\alpha})~(\mathcal{P};\frac{2}{\alpha})(-\mathcal{P};\frac{2}{\alpha}))}\right],
\end{split}
\end{equation}
where $A_{nc}$ and $B_{nc}$ are the constants defined in \cite[Table II]{Alpha_eta_mu}.\\
{\emph{Proof:}} The probability of error for non-fading channel with non-coherent modulation schemes is given as
\begin{equation}\label{Eq:AWGN_Non_Coherent}
P_e(\gamma)=A_{nc} exp(-B_{nc}\gamma).
\end{equation}
Substituting, (\ref{Eq:AWGN_Non_Coherent}) and (\ref{eq:Composite_PDF_SNR_CF}) into (\ref{Eq:BER_Exact}) and using the result proposed in Theorem-6 with $n=0$ and $s=B_{nc}$ in (\ref{eq:MGF_CF}), one will attain (\ref{Eq:BER_Non_Coherent_CF}).\\
\paragraph{Average SEP with Additive Laplacian Noise} The role of Laplacian noise (mean zero and variance $N_0/2$) on wireless signal transmission through indoor, outdoor, and underwater communications has already been noted in the literature \cite{LN1,LN2}. In \cite{LN1} \& \cite{LN3}, error probability analysis for binary phase shift keying (PSK), quadrature PSK, and $M$-ary PSK has been derived over the extended generalized-$K$ and $\alpha$-$\eta$-$\mu$ fading channels. Error rate analysis over previously discussed modulation schemes has been carried out over bivariate Fox's H fading in subsequent corollaries. Also, it is presumed that the receiver is equipped with a minimum Euclidian distance (MED) detector.\\
{\bf{Corollary 3:}} (SEP for Binary and Quadrature PSK). The error probability expressions for binary and quadrature PSK are readily derived, respectively, as
\begin{equation}\label{Eq:BER_BPSK_LN_CF}
\begin{split}
\bar{P}^{BPSK}(e)=\dfrac{\mathcal{R}}{\Gamma(m_s+1)\bar{\gamma}}\mathcal{J}(k_1),
\end{split}
\end{equation}
where $\mathcal{J}(k_1)$ is defined below in which $k_1=2$
\begin{equation}
\mathcal{J}(k_1)= \dfrac{1}{k_1^2}H^{0,2;1,0;1,1}_{2,0;0,1;1,2}\left[^{\frac{\mathcal{T}^{\frac{2}{\alpha}}}{k_1^2\bar{\gamma} m_s}}_{\frac{\hat{\mathcal{T}}^{\frac{\alpha}{2}}}{k_1^2\bar{\gamma} m_s}}\Bigg|^{(-1;2;2)(-m_s,1,1)~~~~-~~~~(\frac{1}{2};\frac{2}{\alpha})}_{~~~-~~~((\hat{\mathcal{P}}-\frac{2}{\alpha});\frac{2}{\alpha})~(\mathcal{P};\frac{2}{\alpha})(-\mathcal{P};\frac{2}{\alpha}))}\right],
\end{equation}
\begin{equation}\label{Eq:BER_QPSK_LN_CF}
\begin{split}
\bar{P}^{QPSK}(e)=\dfrac{3\mathcal{R}}{2\Gamma(m_s+1)\bar{\gamma}}\mathcal{J}(2)+\dfrac{1\mathcal{R}}{4\Gamma(m_s+1)\bar{\gamma}}\\
\times H^{0,2;1,0;1,1}_{2,0;0,1;1,2}\left[^{\frac{\mathcal{T}^{\frac{2}{\alpha}}}{4\bar{\gamma} m_s}}_{\frac{\hat{\mathcal{T}}^{\frac{\alpha}{2}}}{4\bar{\gamma} m_s}}\Bigg|^{(-2;2;2)(-m_s,1,1)~~~~-~~~~(\frac{1}{2};\frac{2}{\alpha})}_{~~~-~~~((\hat{\mathcal{P}}-\frac{2}{\alpha});\frac{2}{\alpha})~(\mathcal{P};\frac{2}{\alpha})(-\mathcal{P};\frac{2}{\alpha}))}\right].
\end{split}
\end{equation}
{\emph{Proof:}} For binary PSK and quadrature PSK, the conditional error probability expressions are, respectively, given as \cite[Eq. (22) \& Eq. (20)]{LN1}
\begin{equation}\label{Eq:LN_BPSK}
P^{BPSK}_e(\gamma)=\dfrac{1}{2} \exp(-2\sqrt{\gamma}),
\end{equation}
\begin{equation}\label{Eq:LN_QPSK}
P^{QPSK}_e(\gamma)=\dfrac{3}{4} \exp(-2\sqrt{\gamma})+ \sqrt{\gamma}\exp(-2\sqrt{\gamma}).
\end{equation}
Plugging (\ref{Eq:LN_BPSK}) and (\ref{eq:Composite_PDF_SNR_CF}) into (\ref{Eq:BER_Exact}) and availing the result provided in Table \ref{table1} completes the proof for BPSK. Similarly, substituting (\ref{Eq:LN_QPSK}) and (\ref{eq:Composite_PDF_SNR_CF}) into (\ref{Eq:BER_Exact}) and availing the result provided in Table \ref{table1} completes the proof for QPSK.\\
{\bf{Corollary 4:}}(SEP for $M$-ary PSK). The closed-form expression for $M$-ary PSK error probability, with $M$$\geq$8, is given as 
\begin{equation}\label{Eq:LN_MPSK_CF}
\bar{P}(e)=\dfrac{8}{M}\sum\limits_{l=0}^{M-1}\mathcal{G}(l)+\dfrac{4tan\big(\frac{\pi}{M}\big)^2}{M\big(1-tan\big(\frac{\pi}{M}\big)^2\big)}\mathcal{J}(k_1),
\end{equation}
where $\mathcal{G}(l)$ is defined as
\begin{equation}\label{Eq:LN_G_L}
\begin{split}
\mathcal{G}(l)=\dfrac{1}{cos\left((2l+1)\frac{2\pi}{M}\right)}
\begin{cases}
cos\left((2l+1)\frac{\pi}{M}\right)^2\mathcal{J}(k_2)\\
-sin\left((2l+1)\frac{\pi}{M}\right)^2\mathcal{J}(k_3)
\end{cases}
\\
-\dfrac{sin\left(\frac{2\pi}{M}\right)}{4\left(cos\left(\frac{2\pi}{M}\right)+sin\left(\frac{4l\pi}{M}\right)\right)}\mathcal{J}(k_4),~~~~~~~~~~~~~~~~~~~~~~~
\end{split}
\end{equation}
where $k_2=\frac{2sin\left(\frac{\pi}{M}\right)}{cos\left((2l+1)\frac{\pi}{M}\right)}$, $k_3=\frac{2sin\left(\frac{\pi}{M}\right)}{sin\left((2l+1)\frac{\pi}{M}\right)}$, and $k_4=2\sqrt{2}cos\left(\frac{2l\pi}{M}-\frac{\pi}{4}\right)$. The values of $\mathcal{J}(k_2)$, $\mathcal{J}(k_3)$, and $\mathcal{J}(k_4)$ are deduced from Table \ref{table1} by replacing the parameter $k$ of the result illustrated at S. No. 1  with $k_2$, $k_3$, and $k_4$, respectively.\\
{\emph{Proof:}} Under Laplacian noise, the conditioned SEP expression for $M$-ary PSK is written as \cite[Eq. (7)]{LN1}
\begin{equation}\label{Eq:LN_MPSK}
{P}(\gamma)=\dfrac{8}{M}\sum\limits_{l=0}^{M-1}\mathcal{G}(l,\gamma)+\dfrac{2tan\big(\frac{\pi}{M}\big)^2}{M\big(1-tan\big(\frac{\pi}{M}\big)^2\big)}\exp(-2\sqrt{\gamma})
\end{equation}
where $\mathcal{G}(l,\gamma)$ is given as
\begin{equation}\label{Eq:LN_G_L_exact}
\begin{split}
\mathcal{G}(l,\gamma)=\dfrac{1}{2cos\left((2l+1)\frac{2\pi}{M}\right)}
\begin{cases}
cos\left((2l+1)\frac{\pi}{M}\right)^2\exp(-\kappa_2\sqrt{\gamma})\\
-sin\left((2l+1)\frac{\pi}{M}\right)^2\exp(-\kappa_3\sqrt{\gamma})
\end{cases}
\\
-\dfrac{sin\left(\frac{2\pi}{M}\right)}{8\left(cos\left(\frac{2\pi}{M}\right)+sin\left(\frac{4l\pi}{M}\right)\right)}\exp(-\kappa_4\sqrt{\gamma}).~~~~~~~~~~~~~~~~~~~~~~~
\end{split}
\end{equation}
Substituting (\ref{Eq:LN_MPSK}) and (\ref{eq:Composite_PDF_SNR_CF}) into (\ref{Eq:BER_Exact}) and with the aid of result from Table \ref{table1}, we evaluate the closed-form expression for $M$-ary PSK, thus concluding the proof.
\subsection{Asymptotic Analysis}
In this subsection, we have incorporated the asymptotic analysis of the previously discussed metrics, as a limiting behaviour in particular environments, i.e., high-power modes. The role of asymptotic analysis is to examine the system behaviour when SNR $\rightarrow \infty$. The analysis helps engineers to optimize the wireless network \cite{Asymptotic}. Among the available approaches, here our analysis is based on the methodology given by Chauhan \emph{et al.} \cite{WLN}. Sticking on the similar approach, the origin PDF for $\alpha$-$\eta$-$\mu$/I-Gamma model is given as 
\begin{equation}\label{Eq. Origin PDF}
f_Y^0(\gamma)\approx \dfrac{\hat{\mathcal{R}}}{\bar{\gamma}^{\alpha\mu}}\gamma^{\alpha \mu-1}+\mathcal{O},
\end{equation}
where $\hat{\mathcal{R}}=\dfrac{\sqrt{\pi}\alpha \mu^{2\mu}h^{\mu}\Gamma(m_s+\alpha \mu)}{\Gamma(\mu)\Gamma(\mu+0.5)\Gamma(m_s)m_s^{\alpha \mu}}$ and $\mathcal{O}$ represents the higher order terms that are being discarded from the actual PDF.
\subsubsection{Outage Probability}
Placing (\ref{Eq. Origin PDF}) into $P_{out}(\gamma_0)=\int\limits_{0}^{\gamma_0}f_Y^0(\gamma) d\gamma$ and utilising simple mathematical manipulations, we obtain
\begin{equation}\label{Eq. Outage_asymptotic}
P_{out}(\gamma_{th})\approx \dfrac{\hat{\mathcal{R}}\gamma_{th}^{\alpha\mu}}{\alpha \mu\bar{\gamma}^{\alpha\mu}}.
\end{equation}
\subsubsection{Average SEP with AWGN}
Substituting (\ref{Eq. Origin PDF}) and (\ref{Eq:AWGN_Coherent}) into (\ref{Eq:BER_Exact}), setting $\sqrt{\mathcal{B}_c\gamma}=r$, and with the inclusion of \cite[Eq. (3.381.4)]{TI}, we obtain the high-power expression for coherent SEP as
\begin{equation}\label{Eq. Asymp_BER_CF}
\bar{P(e)}\approx \dfrac{\hat{\mathcal{R}}\mathcal{A}_c\Gamma\left(\frac{2\alpha \mu+1}{2}\right)}{\sqrt{\pi}\alpha \mu \bar{\gamma}^{\alpha\mu}\mathcal{B}_c^{\alpha \mu}}.
\end{equation}
Similarly, substituting (\ref{Eq. Origin PDF}) and (\ref{Eq:AWGN_Non_Coherent}) into (\ref{Eq:BER_Exact}) and with the aid of \cite[Eq. (3.381.4)]{TI}, yields the solution for non-coherent SEP as
\begin{equation}
\bar{P(e)}\approx \dfrac{\hat{\mathcal{R}}\mathcal{A}_{nc}\Gamma\left(\alpha \mu\right)}{ \bar{\gamma}^{\alpha\mu}\mathcal{B}_c^{\alpha \mu}}.
\end{equation}
\subsubsection{Average SEP with ALN}
Placing (\ref{Eq. Origin PDF}), (\ref{Eq:LN_BPSK}), and (\ref{Eq:LN_QPSK}) into (\ref{Eq:BER_Exact}), setting $2\sqrt{\gamma}=r$ and utilising \cite[Eq. (3.381.4)]{TI}, we obtain the high-power expressions of SEP for BPSK and QPSK, respectively as
\begin{equation}\label{Eq. SEP_LN_BPSK_QPSK}
\bar{P(e)}\approx \dfrac{\mathcal{U}\hat{\mathcal{R}}\Gamma(2\alpha\mu)}{\bar{\gamma}^{\alpha\mu}2^{2\alpha\mu-1}}+\dfrac{\mathcal{V}\hat{\mathcal{R}}\Gamma(2\alpha\mu+1)}{\bar{\gamma}^{\alpha\mu}2^{2\alpha\mu}},
\end{equation}
where for BPSK, $\mathcal{U}=1/2$ \& $\mathcal{V}=0$ and for QPSK, $\mathcal{U}=3/4$ \& $\mathcal{V}=1$.

In similar way, we evaluate the asymptotic result for $M$PSK under ALN by substituting (\ref{Eq. Origin PDF}) and (\ref{Eq:LN_MPSK}) into (\ref{Eq:BER_Exact}), and following the similar methodology as stated previously, we obtain

\begin{eqnarray}\label{Eq. SEP_LN_MPSK}
\begin{split}
\bar{P(e)}\approx \sum\limits_{l=0}^{\frac{M}{4}-1}\mathcal{Z}(l)
+\dfrac{2tan\big(\frac{\pi}{M}\big)^2}{M\big(1-tan\big(\frac{\pi}{M}\big)^2\big)}\dfrac{\hat{\mathcal{R}}\Gamma(2\alpha\mu)}{\bar{\gamma}^{\alpha\mu}2^{2\alpha \mu-1}},
\end{split}
\end{eqnarray}
where
\begin{equation}
\begin{split}
\mathcal{Z}(l)=
\begin{cases}
\dfrac{cos\left((2l+1)\dfrac{\pi}{M}\right)^2}{2cos\left((2l+1)\frac{2\pi}{M}\right)} \dfrac{2\hat{\mathcal{R}}\Gamma(2\alpha\mu)}{\bar{\gamma}^{\alpha\mu}K_1^{2\alpha \mu}}\\
-\dfrac{sin\left((2l+1)\dfrac{\pi}{M}\right)^2}{2cos\left((2l+1)\frac{2\pi}{M}\right)}\dfrac{2\hat{\mathcal{R}}\Gamma(2\alpha\mu)}{\bar{\gamma}^{\alpha\mu}K_2^{2\alpha \mu}}\\
\end{cases}
\\
-\dfrac{sin\left(\frac{2\pi}{M}\right)}{8\left(cos\left(\frac{2\pi}{M}\right)+sin\left(\frac{4l\pi}{M}\right)\right)}\dfrac{2\hat{\mathcal{R}}\Gamma(2\alpha\mu)}{\bar{\gamma}^{\alpha\mu}K_3^{2\alpha \mu}}.\\
\end{split}
\end{equation}
\section{Results and Discussion}\label{Sec: Result}
The proposed analytical frame-works for the bivariate fading model is verified through extensive simulation results plotted in MATLAB software. The section contains mostly graphical visualization for derived metrics and the simulation parameters are well defined in each plot. First, we introduce the plot of power PDFs containing the structure in bivariate Fox's H-function over a practical wireless channel represented through generalised $\alpha$-$\eta$-$\mu$/I-Gamma fading. The plot also contains a few special cases of the distribution that can be achieved by providing suitable parameters. Later, we substantiate other derived results by incorporating the plot for the outage probability metrics. In addition, the preciseness of the presented expressions has been validated by incorporating Monte-Carlo simulation results in all the plots. Throughout the simulation, we use $Format$-1, while $Format$-2 can also be used due to the genericness of the derived expressions. It is also noted that $10^6$ samples are taken for generating the RV for bivariate Fox's H-function over practical wireless channel represented through generalised $\alpha$-$\eta$-$\mu$/I-Gamma fading.

\begin{figure}[!t]
\centering
\includegraphics[width=2.5in]{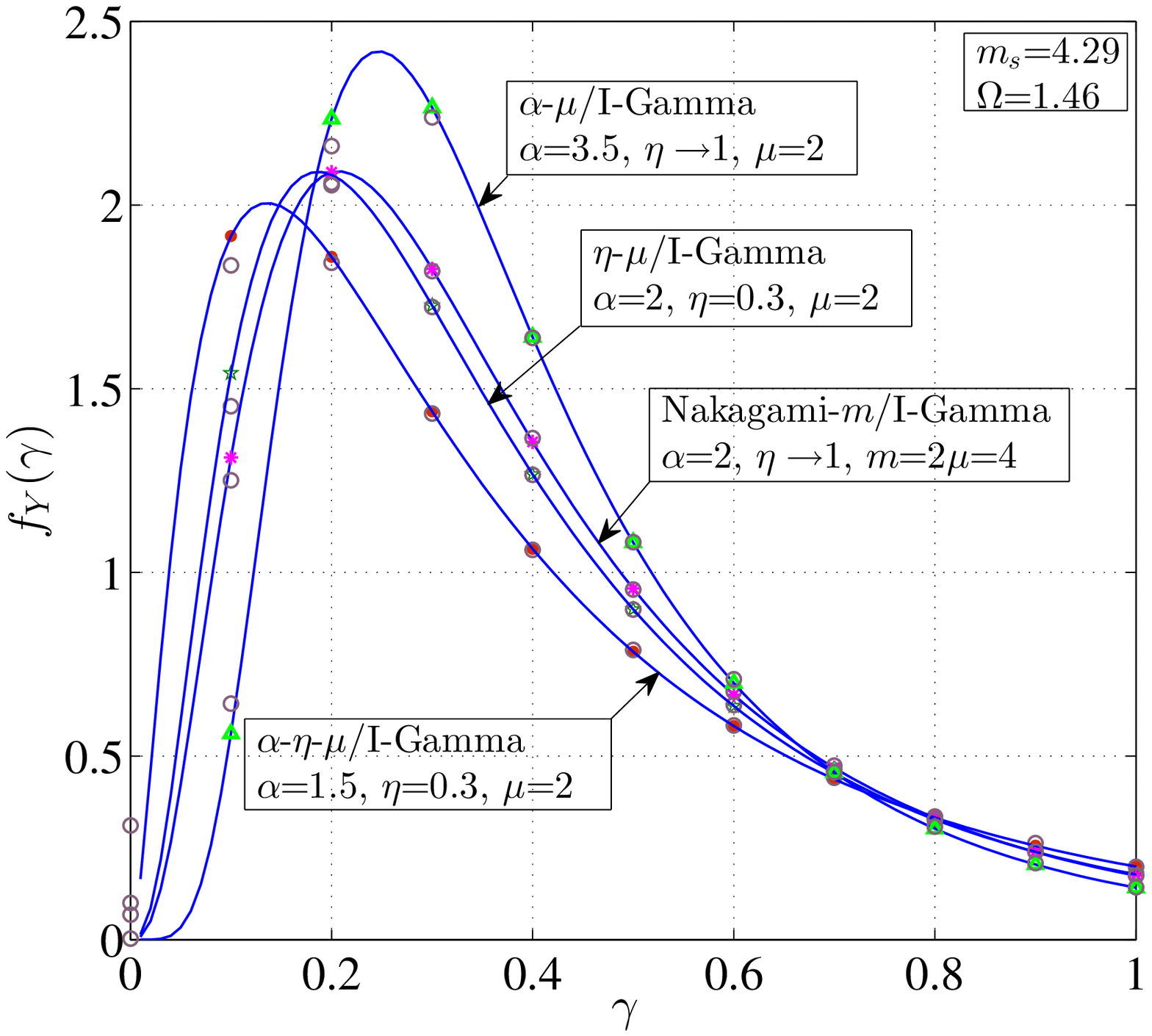}
{\centering(a)}
\includegraphics[width=2.5in]{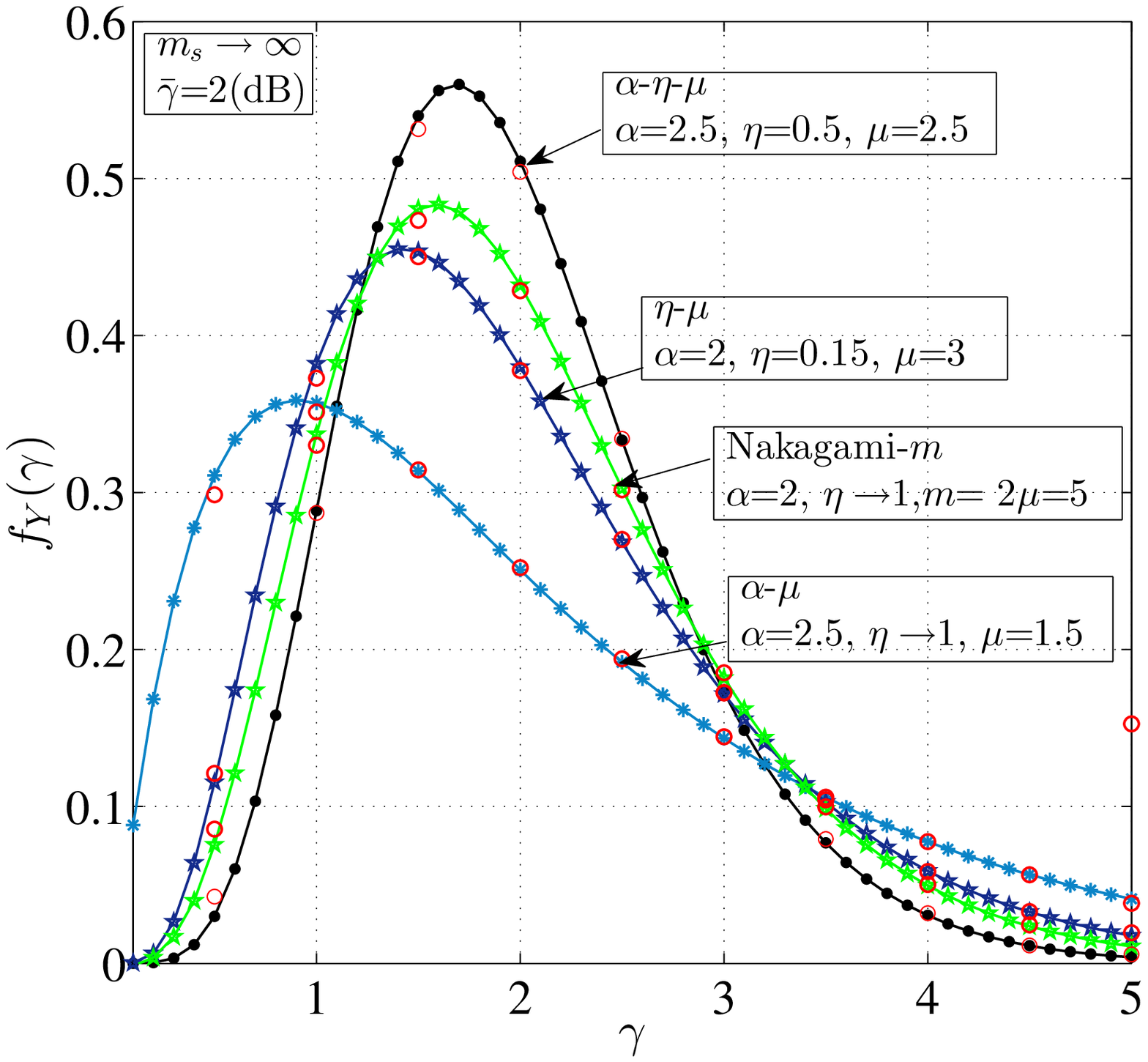}
{\centering{(b)}}
\caption{Power PDFs for special cases of $\alpha$-$\eta$-$\mu$/I-Gamma distribution.}
\label{Fig:Power_PDF_1}
\end{figure}

The power PDFs for composite $\alpha$-$\eta$-$\mu$/I-Gamma distribution and its special cases are demonstrated in Fig. \ref{Fig:Power_PDF_1}. The Figure contains few special cases, such as $\alpha$-$\mu$/I-Gamma, $\eta$-$\mu$/inverse Gamma, Nakagami-$m$/I-Gamma, $\alpha$-$\eta$-$\mu$, $\alpha$-$\mu$, $\eta$-$\mu$, and Nakagami-$m$. The results are further corroborated through Monte-Carlo simulation to substantiate its accurateness, where the results for Monte-Carlo simulations are plotted by directly generating RV for each special case instead of substituting specified values to parameters of $\alpha$-$\eta$-$\mu$/I-Gamma RV.

\begin{figure}[t!]
	\centering
	\includegraphics[width=2.5in]{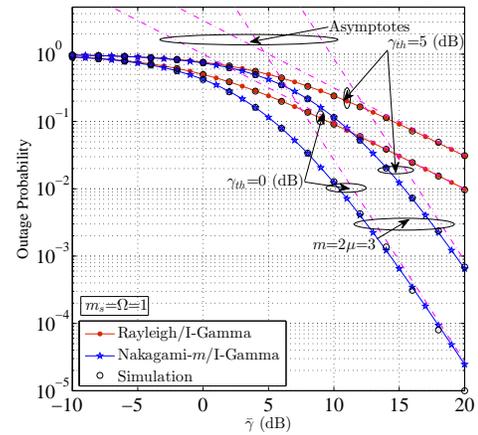}
	\caption{Outage probability versus ${\bar{\gamma}}$ for Rayleigh and Nakagami-$m$ shadowed models.}
	\label{fig: Outage}
\end{figure}

In Fig. \ref{fig: Outage}, system performance with respect to the outage probability metric is analysed. The Figure contains the plot for Rayleigh/I-Gamma and Nakagami-$m$/I-Gamma distribution as special cases of $\alpha$-$\eta$-$\mu$/I-Gamma. The other parameters considered here are $\Omega$=1. $m_s$=1, $\gamma_{th}$=0 \& 5 (dB), and wide range of $\bar{\gamma}$. The Figure also include asymptotic plots for the desired parameters. It is clear from the plot that threshold SNR has an inverse relation with system performance, whereas direct impact is notified against average SNR. It signifies that if the system threshold is made higher then the probability of the signal being discarded by the receiver will be high while it will be lower when average SNR improves. It is further noted from the plots that asymptotic plot merges early for the cases Rayleigh/I-Gamma than Nakagami-$m$/I-Gamma.

\begin{figure}[t!]
	\centering
	\includegraphics[width=2.5in]{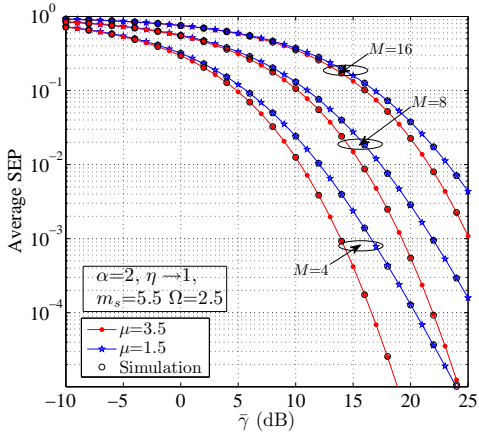}
	\caption{Coherent $M$ary-PSK with varying constellation size.}
	\label{fig: MPSK}
\end{figure}

Average  SEP for coherent $M$-ary PSK is plotted in Fig. \ref{fig: MPSK}, where the impact of constellation size 
on the error probability is portrayed. The channel parameters taken here are, $\alpha$=2, $\eta\rightarrow$1, $\mu$=1.5 \& 3.5, $\Omega$=2.5, $m_s$=2.5. It is quite obvious from the figure that as the constellation size increases the probability of error also increases, resulting in an adverse impact on the system performance as expected. The reduction in SEP is because of reduction in separation between the constellation points. System capability improves as the number of multipath cluster increases due to improvement in the overall received SNR. 

\begin{figure}[t!]
	\centering
	\includegraphics[width=2.5in]{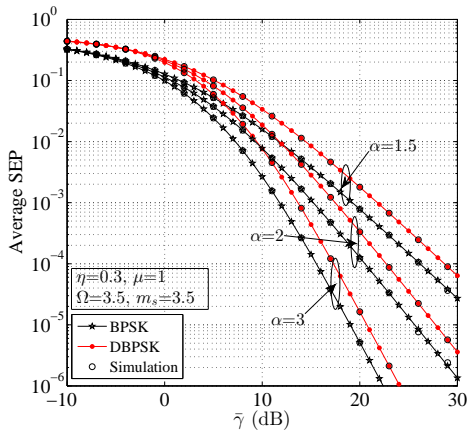}
	\caption{Average SEP for coherent BPSK and non-coherent differential BPSK with different $\alpha$'s.}
	\label{fig:BPSK_DBPSK}
\end{figure}
Average SEP comparison for coherent binary phase-shift keying (BPSK) and non-coherent differential BPSK keying schemes have been plotted in Fig. \ref{fig:BPSK_DBPSK}. The fading parameters are $\eta$=0.3, $\mu$=1, $m_s$=$\Omega$=3.5, and $\alpha$ varies at 1.5, 2, \& 3. It is noted that with an increase in the parameter $\alpha$ and average SNR, error probability reduces. Further, we observe that the error probability for coherent BPSK is less than non-coherent differential BPSK, as expected. It is due to the fact that non-coherent system can not perfectly recover the phase of the transmitted signal that results in a higher error probability relative to a coherent scheme.

\begin{figure}[t!]
	\centering
	\includegraphics[width=2.5in]{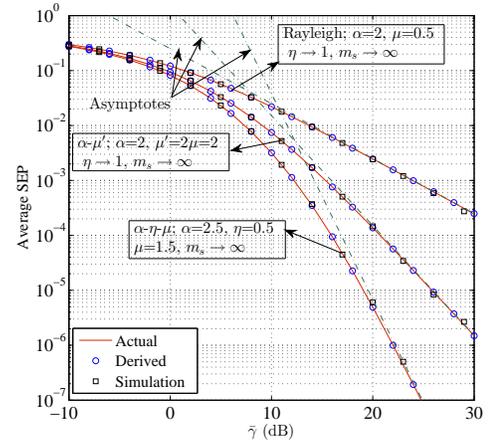}
	\caption{Average SEP for BPSK modulation under Laplacian noise.}
	\label{fig:BPSK_Lapacian}
\end{figure}
Fig. \ref{fig:BPSK_Lapacian} illustrates the plot of bit error rate for BPSK scheme under Laplacian noise. The figure includes actual, derived, asymptotic, and simulation results for Rayleigh, $\alpha$-$\mu$, and $\alpha$-$\eta$-$\mu$ fading models. The actual plot will be generated with the original PDF expression of the respective distributions while derived will be obtained by opting suitable parameters into original BER expression in (\ref{Eq:BER_BPSK_LN_CF}). As expected, the average SEP performance of the channels improves with an increase in the parameters $\alpha$, $\eta$, and $\mu$. The simulation results match well with the analytical plots and verify the accuracy of the proposed formulations.
\section{Conclusion}\label{Sec: Conclusion}
In this correspondence, novel integral and derivative identities were proposed for bivariate Fox's H-function which were helpful in obtaining closed-form statistics of wireless system performance metrics namely, the outage probability, average SEP, and truncated channel inversion with fixed rate capacity. Later, the utility of the derived results are verified through versatile and novel composite $\alpha$-$\eta$-$\mu$/I-Gamma PDF that has taken the analytical structure in bivariate Fox's H-function and derived the results for the outage probability, and average SEP with AWG noise and additive Laplacian noise. The distribution jointly coordinates channel non-linearity and shadowing, it also introduces flexibility to fit in diverse propagation scenarios with that of already proposed in the literature, such as $\eta$-$\mu$/I-Gamma, $\alpha$-$\eta$-$\mu$, Weibull, Nakagami-$m$, etc. In addition, we also derived asymptotic expressions for the above said metrics. The simulation results were included to demonstrate the validity of the presented analytics, where plots of the derived expressions were compared with the Monte-Carlo simulation results. The future aspect of the work will continue in 2nd order statistics and secrecy analysis of the system.

\end{document}